\documentclass[twocolumn,showpacs,prc]{revtex4}
\usepackage{graphicx}
\usepackage{dcolumn}
\usepackage{bm}
\begin{document}

\title{Theory of Metallic Work Functions Between Metals \\ 
and Layers of Exclusion Zone Ordered Water}
\author{A. Widom and J. Swain}
\affiliation{Physics Department, Northeastern University, Boston MA USA}
\author{Y. N. Srivastava}
\affiliation{Physics Department, University of Perugia, Perugia Italy}

\begin{abstract} 
The magnitude of the work function to bring an electron from a metal into 
the exclusion zone water layer making hydrophilic contact with the metallic 
interface is theoretically computed. The agreement with recent experimental  
measurements is satisfactory.  
\end{abstract} 

\pacs{82.47.Jk, 82.47.Uv, 84.60.Jt}
\maketitle

\section{Introduction \label{intro}}

When bulk water is in contact with a hydrophilic metallic 
interface, the water shows an exclusion zone ordered water 
layer that repels large objects such as colloidal particles 
and even heavy ions but can carry extra semi-conducting 
electrons\cite{Zheng:2003,Zheng:2006}. The exclusion zone 
layer of perhaps as thick as hundreds of microns may be 
considered to be an ordered phase of water. Exclusion zone 
layers are sensitive to electromagnetic radiation\cite{Chai:2009} 
at least when in contact with hydrophilic organic substrates. 
The same should hold true for metals. While there are many 
works on the water-metal interface without considerations of the 
layer of exclusion zone water, the understanding how 
electromagnetic radiation with a metallic surface immersed in 
water remains 
incomplete\cite{Thiel:2987,Henderson:2002,
Michaelides:2006,Hodgson:2009,Schnur:2009,Filhol:2007,
Heras:2980,Langenbach:2984}.

The situation changed when experiments\cite{Musumeci:2012} were 
performed yielding data on the work function required to bring 
an electron from the metal into the water. Our purpose is to 
theoretically explain the magnitude of the work function required 
to bring an electron from a metal into the exclusion zone water 
layer making hydrophilic contact with the metallic interface 
The agreement between theory and experimental data is satisfactory.

\section{Polar Water Molecule \label{pw}}

Water molecules in the ideal vapor phase have a static
polarizability\cite{Debye:1928} of the form
\begin{eqnarray}
\alpha_T = \alpha_\infty +\frac{\mu^2}{3k_BT}\ ,
\nonumber \\
\alpha_\infty \approx 1.494\times 10^{-24}\ {\rm cm^3},
\nonumber \\
\mu \approx 1.855\times 10^{-18}\ {\rm Gauss\ cm^3} .
\label{pw1}
\end{eqnarray}
The value of the dipole moment is often incorrectly associated with a
thermal mean dipole moment but this must thought through more carefully 
since in virtue of tumbling motions the thermal dipole moment is null.
The polarizability is given in ideal gas statistical thermodynamics 
\begin{eqnarray}
p_n=e^{(F-E_n)/k_BT} ,
\nonumber \\ 
\alpha_T =-\frac{1}{3}\sum_{nm}
\left[\frac{p_n-p_m}{E_n-E_m}\right] \left|{\bf d}_{nm} \right|^2 ,
\label{pw2}
\end{eqnarray}
wherein \begin{math} p_n\end{math} is the probability of the molecule 
being in state \begin{math} \left|n\right>  \end{math} and
\begin{math} {\bf d}_{nm} \end{math} are the electric dipole matrix 
elements. Define the restricted double sum 
\begin{equation}
\overline{\sum_{nm} }=\sum_{nm} \ \ {\rm restricted\ by\ }
|E_n-E_m|\ll k_BT . 
\label{pw3}
\end{equation}
Eqs.(\ref{pw1}) and (\ref{pw2}) imply 
\begin{equation}
\mu^2 = \overline{\sum_{nm} } 
p_n \left|{\bf d}_{nm}\right|^2
\equiv \overline{|{\bf d}|^2} 
< \left<\left|{\bf d}\right|^2\right>.
\label{pw4}
\end{equation}
with the inequality applying in virtue of the finite but small value
of \begin{math} \alpha_\infty \end{math}. Eq.(\ref{pw4}) defines
the {\em experimental} electric dipole moment 
\begin{math} \mu  \end{math} of a single water molecule given that 
\begin{math} \left<{\bf d}\right>=0 \end{math} due to parity and/or 
time reversal symmetry.

\section{Exclusion Zone Water \label{ezw}}

Metal surfaces tend to be hydrophilic. Water is a liquid 
ferroelectric\cite{Siva:2005}. The mean polarization 
\begin{equation}
{\bf P}({\bf r})=
\left<\sum_k {\bf d}_k \delta({\bf r}-{\bf r}_k)\right>
\label{ezw1}
\end{equation}
is the order parameter for ferroelectricity in water.
The water layer adjacent to the metal is thereby ordered with a 
structure in the polarization 
\begin{math} {\bf P} \end{math} characteristic of a ferroelectric 
boundary. The ferroelectric order is illustrated below in 
FIG. \ref{fig1}. Large objects of micron length scales are excluded 
from the ordered layer since such objects would destroy the 
polarization which would require a positive free energy to destroy 
ferroelectrcity\cite{Gonzalo:1990,Strukov:1997}. Also excluded in 
the ordered water layer are positive ions and some other effectively 
large charged objects. 

What does exists in the ordered exclusion zone  water layer are 
extra electrons described in chemical
terms as \begin{math} OH^- \end{math}. Some protons 
\begin{math} H^+ \end{math} get absorbed into the metal 
leaving behind a net negatively charged exclusion zone layer.
But other extra protons yield a positive charge beyond the water 
metal interface exist above the boundary of the exclusion zone 
layer perhaps a few hundred microns away from the metal 
water interface. The negative charged region which is the exclusion 
zone layer increases with the incidence of electromagnetic radiation 
wherein a battery voltage exists between the positively and negatively 
charged regions within water. 

\begin{figure}
\scalebox {0.5}{\includegraphics{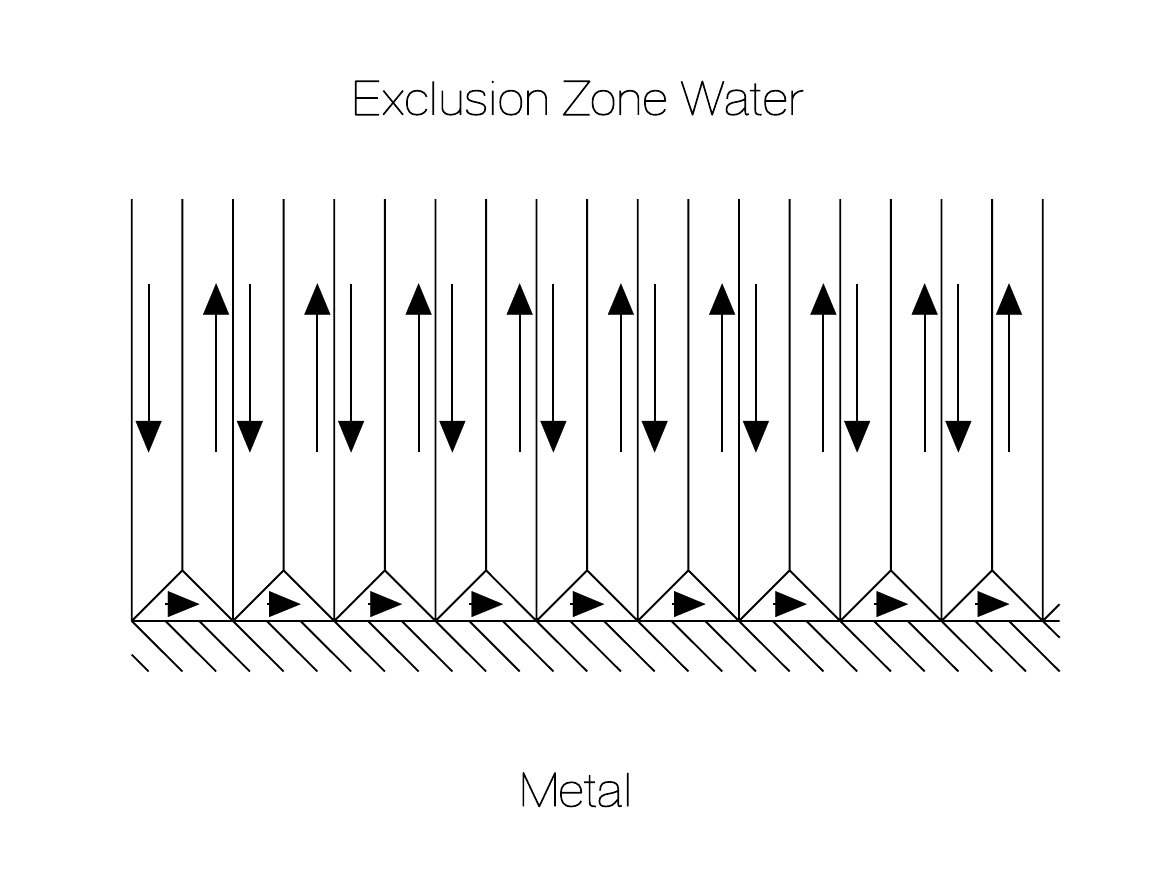}}
\caption{Shown in a schematic fashion is the {\em up-down} 
ordering of the dipole moment per unit volume {\bf P} near 
a boundary of a liquid ferroelectric\cite{Strukov:1997}. 
This is the ordered structure of the exclusion zone 
hydrophilic layer near a liquid ferroelectric water 
metallic interface.}
\label{fig1}
\end{figure}

\section{Work Functions \label{wf}}

\begin{table}
\caption{Work Functions into the Vacuum and into Water}
\label{tab1}
\begin{center}
\begin{tabular}{lrr}
	\hline
		Metal & \ \ \ \ ${\cal W}_{\rm vacuum} $ &\ \ \ \ ${\cal W}_{\rm water}$ \\
	\hline
	\hline
		Pt & \ \ \ \ 5.5 eV &\ \ \ \ 2.1 eV \\
		Au & \ \ \ \ 5.2 eV &\ \ \ \ 2.3 eV \\
		Cu & \ \ \ \ 4.7 eV &\ \ \ \ 2.1 eV \\
	\hline
\end{tabular}
\par\medskip\footnotesize
Experimental determinations\cite{Musumeci:2012} of the work 
${\cal W}_{\rm vacuum} $ required to move an electron from the 
metal into the vacuum and the work ${\cal W}_{\rm water}$ 
required to move an electron from the metal into the exclusion 
zone layer of the water. 
\end{center}
\end{table}

Experimental determinations were discussed\cite{Musumeci:2012} 
for the work required to move an electron from the metal
into the vacuum \begin{math} {\cal W}_{\rm vacuum} \end{math} 
and from the metal into the exclusion zone layer of water. 
\begin{math} {\cal W}_{\rm water} \end{math}. These 
measurements have been listed in TABLE \ref{tab1}.

To remove an electron from an isolated water molecule requires an 
energy of \begin{math} \varphi_0 \approx 12.6\ {\rm eV} \end{math}.
To remove an electron from an isolated water molecule that is initially 
in the first excited state of an isolated water molecule 
requires an energy of 
\begin{math} \varphi_1 \approx 10.2 \ {\rm eV} \end{math}. 
The difference \begin{math} \phi = \varphi_0 - \varphi_1 \end{math} is 
thereby \begin{math} \phi \approx 2.4\ {\rm eV} \end{math}.
As a first approximation, the work function to bring an electron 
from the metal into a ferroelectric ordered layer of water is 
\begin{equation} 
{\cal W}_{\rm water}\approx \phi \approx 2.4\ {\rm eV}.
\label{ebfw1}
\end{equation} 
This work function is in qualitative agreement with data listed 
in TABLE \ref{tab1}. The ordered polarized state of water has 
molecules which are in a coherent superposition of these two 
states\cite{Preparata:1995} with matrix elements of the dipole moment 
\begin{math} {\bf d}_{10} \end{math} which may be taken to be a 
real dipole vector. The state \begin{math} \left|0\right> \end{math} 
is in an electronic {\em s} state forming a scalar while the first 
excited state \begin{math} \left|1\right> \end{math} is in an 
electronic {\em p} state which is thereby triply energy degenerate 
forming an electric dipole vector. This dipole vector may coherently 
rotate within an ordered fluid domain.

\section{Conclusion \label{conc}}

Our purpose was to theoretically explain the magnitude of the work function 
required to bring an electron from a metal into the exclusion zone water 
layer making hydrophilic contact with the metallic interface. Under the 
assumption of quantum coherent polarization ordered 
domains\cite{Siva:2005,Preparata:1995} in the exclusion zone layer, 
the agreement between theory and experimental data is satisfactory.

\section*{Acknowledgments}

J. S. would like to thank the United States National Science Foundation for support under PHY-1205845.

\vfill 
\eject


\begin{thebibliography}{15}

\bibitem{Zheng:2003}
J. Zheng and G.H. Pollack, {\it Phys.Rev.} {\bf E 68},  
3146 (2003).

\bibitem{Zheng:2006}
J. Zheng, W.C. Chin, E. Khijniak, E. Khijniak Jr. and  
G.H. Pollack, {\it Adv. Col. Int. Sci.} {\bf 127}, 9  (2006).  

\bibitem{Chai:2009}
B. Chai, H. Yoo and G.H. Pollack, {\it J. Phys.Chem} {\bf B 113}, 13953 (2009).

\bibitem{Thiel:2987}
P.A. Thiel and T.E. Madey {\it Surf. Sci. Rep.} {\bf 7}, 211 (1987).

\bibitem{Henderson:2002}
M.A. Henderson, 
{\it Surf. Sci. Rep.} {\bf 46}, 1  (2002).

\bibitem{Michaelides:2006}
A. Michaelides, {\it Appl. Phys.} {\bf A 85}, 415  (2006).

\bibitem{Hodgson:2009}
A. Hodgson and S. Haq, {\it Surf. Sci. Rep.} {\bf 64}  381 (2009).

\bibitem{Schnur:2009}
S. Schnur and A. Gro�, {\it New J. Phys.} {\bf 11} 125003 (2009).

\bibitem{Filhol:2007}
J.S. Filhol and M.L.Bocquet, {\it Chem. Phys. Lett.} {\bf 438}, 203 (2007).

\bibitem{Heras:2980}
J.M.Heras and L.Viscido, {\it Appl. Surf. Sci.} {\bf 4}, 238 (1980.

\bibitem{Langenbach:2984}
]E. Langenbach, A. Spitzer, H. Luth, {\it Surf.Sci.} {\bf 147}, 
179 (1984).

\bibitem{Musumeci:2012}
F. Musumeci and G.H. Pollack, {\it Chem. Phys. Lett.} 
{\bf 536}, 65  (2012). 

\bibitem{Debye:1928}
P. Debye, ``Polar Molecules'' p 30, Dover Publications Inc.,
New York (1928).

\bibitem{Siva:2005}
S. Sivasubramanian, A. Widom and Y.N. Srivastava,
{\it Physica} {\bf A 345}, 356 (2005).

\bibitem{Gonzalo:1990}
J.A. Gonzalo, ``Effective Field Approach to Phase Transitions 
and Some Applications to Ferroelecrics'', World Scientific 
Singapore (1990).

\bibitem{Strukov:1997}
B.A. Strukov and A.P. Levanyuk, 
``Ferroelecric Phenomena in Crystals'', Chapt. 10, p 219, 
Springer  Berlin (1997).

\bibitem{Preparata:1995}
G. Preparata, ``QED Coherence in Matter'', Chapt. 10,
World Scientific Singapore (1995).


\end{thebibliography}
\end{document}